\documentclass[aps,nofootinbib,preprintnumbers,twocolumn,superscriptaddress,prd]{revtex4-1}

\usepackage{amsmath}
\usepackage{amssymb}
\usepackage{slashed}
\usepackage{epsfig}
\usepackage{bbm,bm}
\usepackage{hyperref}
\usepackage{color}
\usepackage{comment}
\usepackage{amsfonts}
\usepackage{dsfont}

\def\bea{\begin{eqnarray}} 
\def\eea{\end{eqnarray}}
\def\be{\begin{equation}} 
\def\ee{\end{equation}} 
\def\ba{\begin{array}}
\def\ea{\end{array}} 
\def\nn{\nonumber}

\def\be{\begin{equation}}
\def\ee{\end{equation}}
\def\bea{\begin{eqnarray}}
\def\eea{\end{eqnarray}}

\def\nn{\nonumber}

\begin{document}

\title{Multi-critical $\square^k$ scalar theories: A perturbative RG approach with $\epsilon$-expansion}

\author{M.\ Safari}
\email{safari@bo.infn.it}
\affiliation{INFN - Sezione di Bologna, via Irnerio 46, 40126 Bologna, Italy}
\affiliation{
Dipartimento di Fisica e Astronomia,
via Irnerio 46, 40126 Bologna, Italy}

\author{G.\ P.\ Vacca}
\email{vacca@bo.infn.it}
\affiliation{INFN - Sezione di Bologna, via Irnerio 46, 40126 Bologna, Italy}

\begin{abstract}
We employ perturbative RG and $\epsilon$-expansion to study multicritical single-scalar field theories with higher derivative kinetic terms of the form $\phi(-\Box)^k\phi$. 
We focus on those with a $\mathbb{Z}_2$-symmetric critical point which are characterized by an upper critical dimension $d_c=2 n k/(n-1)$ accumulating at even integers. 
We distinguish two types of theories depending on whether or not the numbers $k$ and $n-1$ are relatively prime.
When they are, the critical theory involves a marginal power-like interaction $\phi^{2n}$ and the deformations admit a 
derivative expansion that at leading order involves only the potential. In this case we present the beta functional of the potential
and use this to calculate some anomalous dimensions and OPE coefficients. These confirm some CFT data obtained using conformal-block techniques, while giving new results. In the second case where $k$ and $n-1$ have a common divisor, the theories show a much richer structure induced by the presence of marginal derivative operators at criticality. We study the case $k=2$ with odd values of $n$, which fall in the second class, and calculate the functional flows and spectrum. These theories have a phase diagram characterized at leading order in $\epsilon$ by four fixed points which apart from the Gaussian UV fixed point include an IR fixed point with a purely derivative interaction.

\end{abstract}

\maketitle

\section{Introduction}

Universal large distance behavior of many physical systems are well described within the theoretical framework of quantum/statistical field theory at criticality. A key notion is the renormalization group (RG) which has brought much insight into the concept of universality as well as providing a groundwork to calculate universal quantities. The fixed points of the RG flow correspond to scale invariant theories which are relevant for addressing critical phenomena. 

While fundamental physical theories may require unitarity as an essential ingredient, non-unitary critical theories have also found interesting applications in physics. 
Among them is the well studied Lee-Yang edge singularity which in a Langau-Ginzburg (LG) description is associated to a $\phi^3$ potential~\cite{Fisher:1978pf}.
Further, there are the so called higher derivative theories, applicable for instance to the theory of elasticity~\cite{Nakayama:2016dby} and to particular quartic-derivative models which describe the isotropic phase of critical Lifshitz theories, and may be relevant for the physics of certain polymers~\cite{polimers}. 
The latter has been also studied with $\epsilon$-expansion techniques~\cite{Hornreich:1975zz,diehl} and recently also with non perturbative functional RG methods~\cite{Bonanno:2014yia}.  

On the other hand scale invariant theories, whether unitary or not, are often also conformal. Moreover there is growing evidence that when accompanied by unitarity, scale invariance is enhanced to conformal invariance. In particular, this is long proven to be true in two dimensions and established more recently at least in a perturbative setting, also in four dimensions.   
The presence 
of conformal symmetry in most scale invariant theories of interest 
further motivates the vast amount of research which has been devoted in recent years to studying critical systems 
relying solely on conformal symmetry and without referring to RG~\cite{ElShowk:2012ht,Gliozzi:2016ysv,Gliozzi:2017hni,Alday:2016njk,Rychkov:2015naa,Basu:2015gpa,Nii:2016lpa,Codello:2017qek}. 

Apart from physical applications, studying non-unitary critical theories and in particular conformal field theories (CFT) especially in higher dimensions is interesting from a theoretical point of view 
as it improves our 
understanding of the structure of quantum and statistical field theories.

Other examples of non-unitary critical theories studied in the literature include the whole family of odd-potential multicritical theories \cite{Codello:2017epp} generalizing the Lee-Yang edge singularity. In particular a class of non-unitary theories which are $\phi^{2n}$ deformations of generalized free CFTs with higher derivative kinetic term of the form  $\phi (-\Box)^k \phi$ have recently been investigated in~\cite{Gliozzi:2016ysv,Gliozzi:2017hni} using the structure of conformal blocks. The corresponding free theories had been earlier analysed in \cite{Brust:2016gjy} mainly motivated by de Sitter holography. Theories with higher derivative kinetic terms in higher dimensions have been considered also in \cite{Osborn:2016bev} and from a pure RG perspective in \cite{Gracey:2017erc}.

Inspired by these recent analyses of non-unitary CFTs we investigate further such models with a higher derivative kinetic term starting from a consistent LG description. The pathologies of higher derivative scalar theories brought up in \cite{Aglietti:2016pwz} do not apply here as there is no explicit scale present in the theory. We mainly employ RG techniques but as crosscheck we also make few analysis using the Schwinger-Dyson equation (SDE) and CFT constraints on the correlation functions. Both methods have already been applied to theories with a standard kinetic term \cite{Nii:2016lpa, Codello:2017qek, ODwyer:2007brp, Codello:2017hhh, Codello:2017epp}. The picture that emerges in the general case has a much richer structure than that previously unveiled. 

In particular we find that theories with $k>1$ admit the simple LG description of generalized Wilson-Fisher type with $\phi^{2n}$ critical interaction only when $n-1$ and $k$ are coprime numbers. In this case we confirm the anomalous dimensions, pushing the computation to order $\epsilon^2$ for the relevant operators, and calculate an infinite family of operator product expansion (OPE) coefficients, a finite number of whom which have so far been accessible with other methods coincide with these earlier results. 

Instead when $n-1$ and $k$ have a common divisor the critical theories are more involved. 
They are characterized by the presence of marginal derivative interactions of order less than $2k$. In such cases one expects, for fixed $k$ and $n$, several non trivial critical solutions. This is confirmed in a detailed analysis we make for the case $k=2$ and $n=2m+1$. We report several anomalous dimensions for such cases and have a first look at the phase diagram which is given for the representative case $n=3$.
Here we present the main results and postpone most of the computational details to a separate work~\cite{Safari:2017tgs}.

\section{RG approach to $\Box^k$ theories: general considerations} \label{sec.general}
In this work we are interested in using perturbation theory to investigate critical theories which are smooth deformations of the generalized free CFT 
\be 
\mathcal{L}_{F} = {\textstyle{\frac{1}{2}}}\phi\, (-\square)^k \phi.
\ee
We concentrate on those theories with upper critical dimension $d_c=2 n k/(n-1)$, which can be fixed by the requirement that $\phi^{2n}$ be a marginal operator. This also fixes the critical dimension of the field to $\delta_c = k/(n-1)$. We restrict here to integer values of $n$. 

In order to identify possible scale invariant deformations it is useful to determine all marginal operators. By dimensional analysis, an operator with $2l$ number of derivatives and $p$ number of fields is marginal if the following condition holds
\be 
\frac{2n-p}{n-1}k=2l.
\ee
From this simple relation it is clear that when $n-1$ and $k$ are relatively prime, and when $k$ is odd, then (apart from the kinetic operator) $\phi^{2n}$ is the only marginal operator present. For even values of $k$ instead there exist other marginal operators with $k$ number of derivatives and $n+1$ number of fields which is an odd number, but these can be consistently omitted because of their odd $\mathbb{Z}_2$ parity, and this is indeed what we do in the present work. 

On the other hand, when $n-1$ and $k$ have a common divisor, there always exist marginal derivative-operators with even parity. The beta function of their corresponding couplings include terms with pure powers of the $\phi^{2n}$-coupling, and therefore such flows will only lead to non-trivial fixed points which are either pure derivative-interactions or a mixture of derivative and potential-interactions. Consequently these operators cannot be avoided and the local potential approximation breaks down. 


We refer to theories with coprime $k$ and $n-1$ as ``first type'' and otherwise as ``second type''. Based on this classification, in the following we consider these two type of theories in turn and perform an $\epsilon$-expansion by moving away from the critical dimension to $d=d_c-\epsilon$ and employ perturbative RG in the functional form. 

\section{Theories of the first type} \label{sec.type1}

We concentrate in this section on theories of the first type where $k$ and $n-1$ are relatively prime. In this case one can consistently consider pure potential deformations
\be 
\mathcal{L} = {\textstyle{\frac{1}{2}}}\phi\, (-\square)^k \phi + V(\phi)\,,
\ee
where the field $\phi$ implicitly contains the wave-function so that the kinetic term is always in the canonical form. In order to extract critical information we express variables in units of the RG scale $\mu$
\be
v(\varphi) = \mu^{\!-d}\, V(\mu^\delta\,\varphi)\,,
\ee
where $\delta = d/2-k$ is the dimension of the field. The beta functional of the dimensionless potential at cubic order in the couplings is found to be 
{\setlength\arraycolsep{2pt}
\bea \label{betav}
\displaystyle \beta_v &=&  -\,d v + \frac{d- 2k+\eta}{2} \varphi \, v'+ \frac{1}{n!}\, v^{(n)\,2}  \nn\\
&& -\Gamma(n\delta_c)\,\frac{1}{3}\sum_{r,s,t}\frac{K^{n,k}_{rst}}{r!s!t!} \;v^{(r+s)}v^{(s+t)}v^{(t+r)}  \nn\\
&& -\frac{1}{n!} \sum_{s,t}\!\frac{J^{n,k}_{st}}{s!t!}\, v^{(n)}v^{(n+s)}v^{(n+t)}\,,
\eea}%
where, in order to simplify the expression, and as will become clear later, accord with the CFT normalization, we have found it convenient to rescale the potential as 
\be 
v\rightarrow v\;[(4\pi)^k\Gamma(k)]^n \;\Gamma(n\delta_c)/\Gamma(\delta_c)^n.
\ee
The sums in \eqref{betav} run over positive integers such that $r+s+t=2n$ and $r,s,t\neq n$ in the first sum and $s+t=n$ in the second sum. Also, we have defined
\bea 
K^{n,k}_{rst} &\equiv& \frac{\Gamma\left((n-r)\delta_c\right)\Gamma\left((n-s)\delta_c\right)\Gamma\left((n-t)\delta_c\right)}{\Gamma\left(r\delta_c\right)\Gamma\left(s\delta_c\right)\Gamma\left(t\delta_c\right)},\\
J^{n,k}_{st} &\equiv& \psi(n\delta_c) - \psi(s\delta_c)-\psi(t\delta_c) + \psi(1)\,,
\eea
where $\psi(x)=\Gamma'(x)/\Gamma(x)$ is the digamma function. The quadratic term in \eqref{betav} comes from a diagram with $n-1$ loops, while the cubic term appears at $2(n-1)$-loop order. The potential induces a flow for the kinetic term coefficient which can be used to fix the anomalous dimension $\eta$ in \eqref{betav}. In terms of the marginal coupling $g \equiv v^{(2n)}(0)/(2n)!$ this is 
\be \label{eta}
\eta = (-1)^{k+1}\,
\frac{n(\delta_c)_k}{k(\delta_c+k)_k}\;4(2n)!\, g^2 \,,
\ee
where  $(a)_b = \Gamma(a+b)/\Gamma(a)$ is the Pochhammer symbol. From the general structure of the beta function \eqref{betav} it is clear that the coefficient of the quadratic term $g^2$ is positive. This ensures the presence of a Wilson-Fisher type fixed point. It is a straightforward task to find the value of this fixed point at quadratic order in $\epsilon$ 
{\setlength\arraycolsep{2pt}
\bea 
\frac{(2n)!^2}{n!^3} g &=& \displaystyle  (n-1)\epsilon - n \,\eta   \\ 
&& \hspace{-15mm}+ \frac{n!^4(n-1)^2}{(2n)!}\bigg[\frac{n!^2}{3}\,\Gamma(n\delta_c)\sum_{r,s,t}\frac{K^{n,k}_{rst}}{(r!s!t!)^2} +\sum_{s,t}\frac{J^{n,k}_{st}}{(s!t!)^2} \bigg]\epsilon^2 \nn .
\eea}%
where here $\eta$ is expressed in terms of $\epsilon$ adopting the leading $\epsilon$ dependence of the fixed point coupling
\be \label{etaep}
\eta =  (-1)^{k+1}\,
\frac{n(\delta_n)_k}{k(\delta_n+k)_k}\; \frac{4(n-1)^2n!^6}{(2n)!^3}\,\epsilon^2.
\ee
Using the value of the fixed point and the stability matrix of the flow \eqref{betav} one calculates the critical exponents at second order in $\epsilon$
{\setlength\arraycolsep{2pt}
\bea  \label{cad} 
&& \tilde\gamma_i = \displaystyle \frac{\eta}{2} i + 2n\, \eta\, \delta^{2n}_i 
+ \frac{(n-1)i!}{(i-n)!}\,\frac{2n!}{(2n)!}\left[\epsilon - \frac{n}{n-1} \,\eta\right] \nn\\
&& + \frac{(n\!-\!1)^2i!n!^6}{(2n)!^2}\Gamma(n\delta_c)\!\sum_{r,s,t}\!\frac{K^{n,k}_{rst}}{(r!s!t!)^2} \!\left[\!\frac{2n!}{3(i\!-\!n)!} \!-\!\frac{r!}{(i\!-\!2n\!+\!r)!}\!\right]\!\epsilon^2 \nn\\
&& +\frac{(n-1)^2i! n!^4}{(2n)!^2}\sum_{s,t}\frac{J^{n,k}_{st}}{s!^2t!^2}\left[\frac{n!}{(i\!-\!n)!} -\frac{2s!}{(i\!-\!2n\!+\!s)!}\right]\!\epsilon^2 .
\eea}%
This order-$\epsilon^2$ expression is valid for the relevant and marginal operators, while at order $\epsilon$ its range of validity extends to all the irrelevant operators as well. These leading order results for $\tilde\gamma_i$ are also independent of $k$.

The order-$\epsilon$ expressions for the critical exponents \eqref{cad} as well as the anomalous dimension \eqref{etaep} are in agreement with \cite{Gliozzi:2016ysv,Gliozzi:2017hni}. Moreover, independent of $k$ the operator $\phi^2$ is always relevant and therefore its corresponding critical exponent $\gamma_2$ at second order in $\epsilon$ is read off from \eqref{cad} for any value of $k,n$. In particular for $n>2$ it takes a simple form with no order-$\epsilon$ piece
\be 
\tilde\gamma_2 \!= \!\left[\!\frac{(-1)^{k+1} n(\delta_c)_k}{k(\delta_c\!+\!k)_k}  
\!-\!\frac{n(2n\!-\!1)\Gamma(k)(\delta_c)_k}{(k)_k(\delta_c\!-\!k)_k}\!\right]\!\!\frac{4(n\!-\!1)^2n!^6}{(2n)!^3}\,\epsilon^2
\ee
which for $k=1$ reproduces the result of \cite{Codello:2017qek}. Finally, expanding the beta function \eqref{betav} around the fixed point, a family of OPE coefficients can be obtained by extracting the coefficient of the quadratic terms in the couplings
{\setlength\arraycolsep{2pt}
\bea 
&& \tilde C^l_{\;ij}\! = \frac{1}{n!}  \frac{i!}{(i\!-\!n)!}\frac{j!}{(j\!-\!n)!} \!+\!2(2n)!\left(i\delta^{2n}_j \!\!+\! j\delta^{2n}_i\!\!+\!2n\,\delta^{2n}_i\delta^{2n}_j\right)\!\epsilon \nn\\
&& -\Gamma(n\delta_n) \frac{(n-1)n!^3}{(2n)!}\sum_{r,s,t}\frac{K^{n,k}_{rst}}{r!s!t!^2} \,\frac{j!}{(j-s-t)!}\frac{i!}{(i+s-2n)!}\,\epsilon \nn\\
&& - \frac{(n-1)n!^2}{(2n)!}\sum_{s,t}\frac{J^{n,k}_{st}}{s!t!} \;\left[\frac{1}{n!}\frac{j!}{(j-n-s)!}\frac{i!}{(i-n-t)!} \right. \nn\\
&& +\! \left.\frac{1}{s!}\frac{i!}{(i-n)!}\frac{j!}{(j-n-s)!} + \frac{1}{s!}\frac{j!}{(j-n)!}\frac{i!}{(i-n-s)!}\right]\!\epsilon 
\eea}%
These are the dimensionless OPE coefficients with $l=i+j-2n$. The subset of these OPE coefficients with $l=1$, which can be parametrized by a single parameter $m=1,\cdots,n-1$, takes a significantly simplified form
{\setlength\arraycolsep{2pt}
\bea  
\tilde C^1_{\;n-m, n+m+1} &=& \\
&& \hspace{-25mm} - \frac{\left(\delta_c\right)_k\,(n-1)\Gamma\left(k\right)}{\left(-m\delta_c\right)_k\left((m+1)\delta_c\right)_k}\,\frac{(n+m+1)!(n-m)!}{(n-m-1)!(n+m)!} \,\frac{n!^3}{(2n)!}\,\epsilon \nn
\eea}%
which matches the dimensionless subset of OPE coefficients in \cite{Gliozzi:2017hni} when the difference in the normalization of the fields is taken into account.

\noindent
{\bf \it CFT approach.}
We have already seen that with our RG-based analysis one can obtain
some results previously derived at leading order in $\epsilon$ using a conformal block analysis in~\cite{Gliozzi:2016ysv,Gliozzi:2017hni}.
Alternatively, again assuming that the critical theory is a CFT, we can use the constraints from conformal invariance on the two and three point correlation functions along with the SDE to extract some critical properties. Here we illustrate only the procedure. More details will be given elsewhere~\cite{Safari:2017tgs}. 
We define the scaling dimensions of the composite operators $\phi^i$ as $\Delta_i=(k/(n-1)-\epsilon/2)i +\gamma_i$, where $\gamma_i$ are the anomalous dimensions.
Closely following \cite{Codello:2017qek} which applies this idea to the case $k=1$, one can reproduce $\eta$ \eqref{eta} by applying $\Box^k_x\Box^k_y$ to the two point function $\left\langle \phi_x\phi_y\right\rangle$, both directly and by using the equation of motion $\Box^k\phi = 2 n g (-1)^{k-1} \phi^{2n-1}$. 
The latter leads to a new correlator which, for a leading order analysis, can be computed in the free theory. 
Similarly applying $\Box^k_x$ to $\left\langle \phi_x\phi^i_y\phi^{i+1}_z\right\rangle$ and using the SDE gives a recursion relation for the $\gamma_i$ which gives the order-$\epsilon$ values of the critical exponents \eqref{cad} upon using the boundary condition $\gamma_{n-1} = \mathcal{O}(\epsilon^2)$. Finally applying $\Box^k_x$ to the three point function $\left\langle \phi_x\phi^{2k}_y\phi^{2l+1}_z\right\rangle$
one can obtain the structure constants $C_{1,2k,2l+1}$ at order $O(\epsilon)$, which are compatible with our RG analysis and coincide with the results of~\cite{Gliozzi:2016ysv,Gliozzi:2017hni}, if the $\epsilon$-dependence of the critical coupling $g(\epsilon)$ is used. This can be derived from the relation $\Delta_{2n-1}=2k+\Delta_1$.

\section{Theories of the second type: The case $k=2$}
%
We now move to $\Box^k$ multicritical theories where $k$ and $n-1$ have a non-trivial common divisor. 
As anticipated earlier, this case is considerably more involved than the first class. 
In order not to complicate the calculations and at the same time convey the qualitative features of such theories we stick to the case $k=2$. 
The case of even $n$ falls in the first class already discussed, so in this section we take $n=2m+1$. The upper critical dimension and the field dimension at criticality are
\bea 
d_c = 4 + 2/m, \qquad \delta_c = 1/m\,.
\eea
Notice that for $m=1,2$ the upper critical dimension is an integer. The marginal operators for general $m$ are $\phi^{2(2m+1)}$ and $\phi^{2m}(\partial\phi)^2$, apart from the kinetic operator. As argued before one is forced in this case to take into account 2-derivative operators as well. We therefore consider the following Lagrangian
\bea
\mathcal{L} = {\textstyle{\frac{1}{2}}}\phi\,\square^2 \phi +{\textstyle{\frac{1}{2}}}Z(\phi)(\partial\phi)^2+ V(\phi)\,.
\label{LGnew}
\eea
In terms of the dimensionless field $\varphi$ the dimensionless functions ($d=d_c\!-\!\epsilon$ and $\delta = d/2\!-\!2$) are defined as 
\bea
v(\varphi) = \mu^{\!-d}\, V(\mu^\delta\,\varphi), \qquad z(\varphi) = \mu^{\!-2}\, Z(\mu^\delta\,\varphi)\,,
\eea
and their flow equations at quadratic order are
\bea \label{bv}
\beta_v &=& -d v \!+\! \frac{d\!-\! 4}{2} \varphi v'+ \frac{v^{(m\!+\!1)}z^{(m\!-\!1)}}{(m\!+\!1)!} + \frac{(v^{(2m\!+\!1)})^2}{(2m\!+\!1)!} \,,\\
%
%
\label{bz}
\beta_z &=& -2 z \!+\! \frac{d\!-\! 4}{2} \varphi  z'\!+\! 2 \frac{v^{(2m+1)}z^{(2m+1)}}{(2m\!+\!1)!}  +\frac{z^{(m+1)}z^{(m-1)}}{(m\!+\!1)!} \nn\\
&& +  \frac{3m\!+\!2}{2(2m\!+\!1)}\,\frac{(z^{(m)})^2}{(m+1)!} -\frac{2(m+1)}{(2m+1)}\,\frac{(v^{(3m+2)})^2}{(3m\!+\!1)!}\,,
\eea
where the following rescalings have been done
\be \label{res2}
v\rightarrow (m+1)\,\frac{(4\pi)^{2(2m+1)}}{m^2\Gamma^{2m}(\delta_c)}\, v, \qquad
z\rightarrow \frac{(4\pi)^{2m+1}}{m\Gamma^{m}(\delta_c)}\, z.
\ee
The anomalous dimension $\eta$ has been dropped from the linear term as it does not contribute at this level of approximation. The quadratic terms in the beta functions come from melon type diagrams which are of $m$-loop order for the $v^{(i)}z^{(j)}$ term in \eqref{bv} and of $2m$-loop order for the $(v^{(i)})^2$ term. 
Also the $v^{(i)}z^{(j)}$ term in \eqref{bz} appears at $2m$-loops, the $z^{(i)}z^{(j)}$ term at $m$-loops and the $(v^{(i)})^2$ term at $3m$-loops. 
The functions $Z(\phi)$ and $V(\phi)$ induce a flow on the coefficient of the kinetic term which is implicit in the definition of the field $\phi$. 
This can be used to fix the anomalous dimension. In terms of the two couplings $g,h$ which are respectively the coefficients of the operator $\varphi^{2(2m+1)}$ in $v(\varphi)$ and $\varphi^{2m}(\partial\varphi)^2$ in $z(\varphi)$, the anomalous dimension is 
{\setlength\arraycolsep{2pt}
\bea  \label{eta2}
\eta &=& \frac{\Gamma(\delta_c)}{m^2\Gamma(2+\delta_c)}\, \frac{(2m)!}{2(3m+1)}h^2 \\
&-& \frac{\Gamma(\delta_c)}{m^4\Gamma(4+\delta_c)}4(m+1)^2(2m+1)^2(4m+1)!\,g^2 \,.\nn
\eea}%
Also from an analysis of the stability matrix, one can calculate the critical exponents of the relevant and marginal operators. 
Although straightforward to calculate, here we report only those that are not affected by mixing, that is, the ones for which the stability matrix is either diagonal or lower triangular. For the potential operators these are given by
\be \label{cad2}
\tilde\gamma_i = \frac{i!}{(i\!-\!m\!-\!1)!}\,\frac{(2m)!}{(m\!+\!1)!^2}h + \frac{i!}{(i\!-\!2m\!-\!1)!}\frac{2(2(2m\!+\!1))!}{(2m\!+\!1)!^2}g,
\ee 
where $g,h$ are assumed to be at the critical point. This equation is valid for $i=0,\cdots, 3m+1$, bearing in mind that terms with negative factorials in the denominator are interpreted as zero. In particular for $i\leq m$ the critical exponents $\tilde\gamma_i$ vanish at linear order in the couplings.

\begin{figure}[h]
\includegraphics[width=42.7mm]{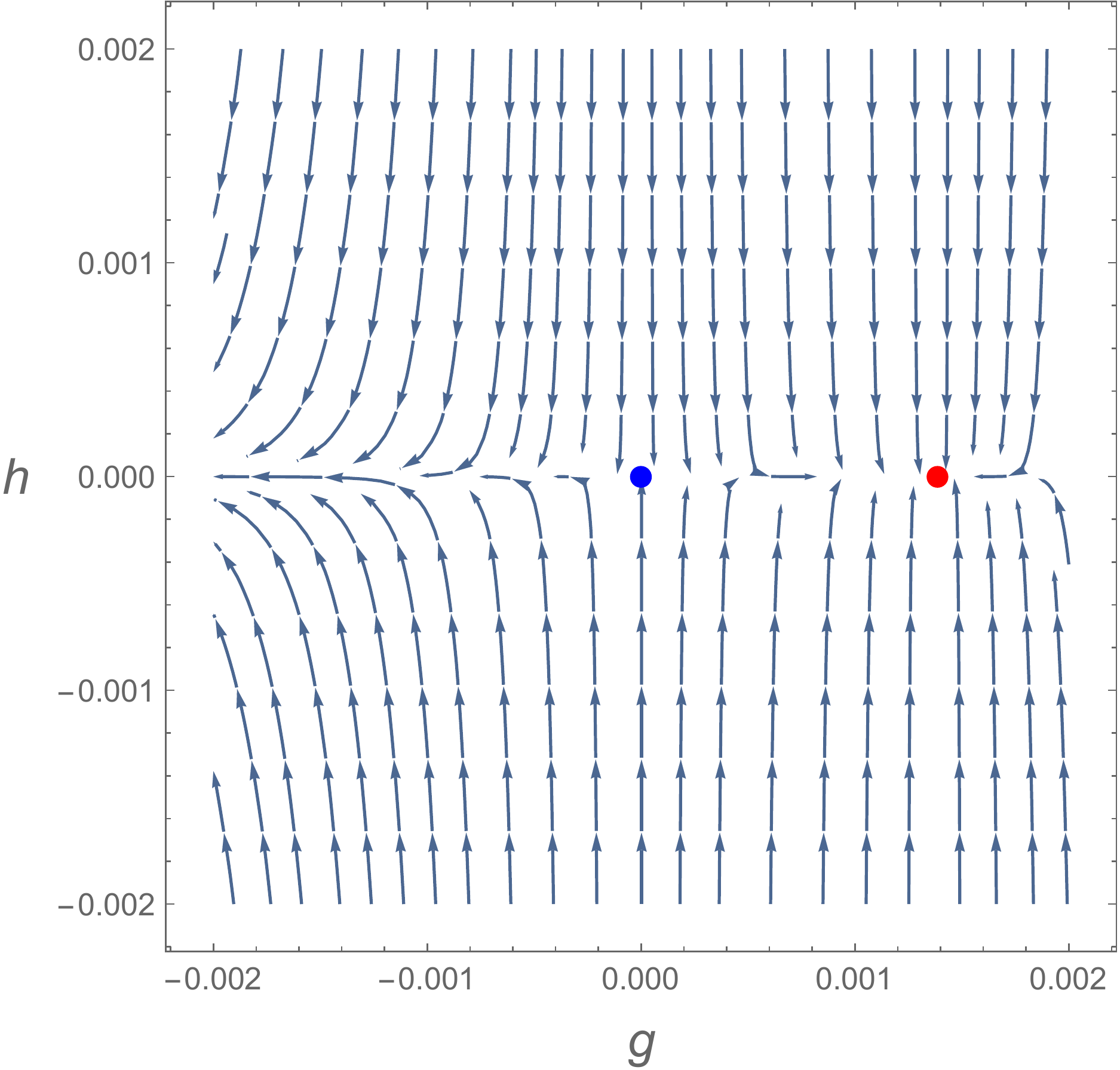}
\hspace{0.0cm}
\includegraphics[width=41.6mm]{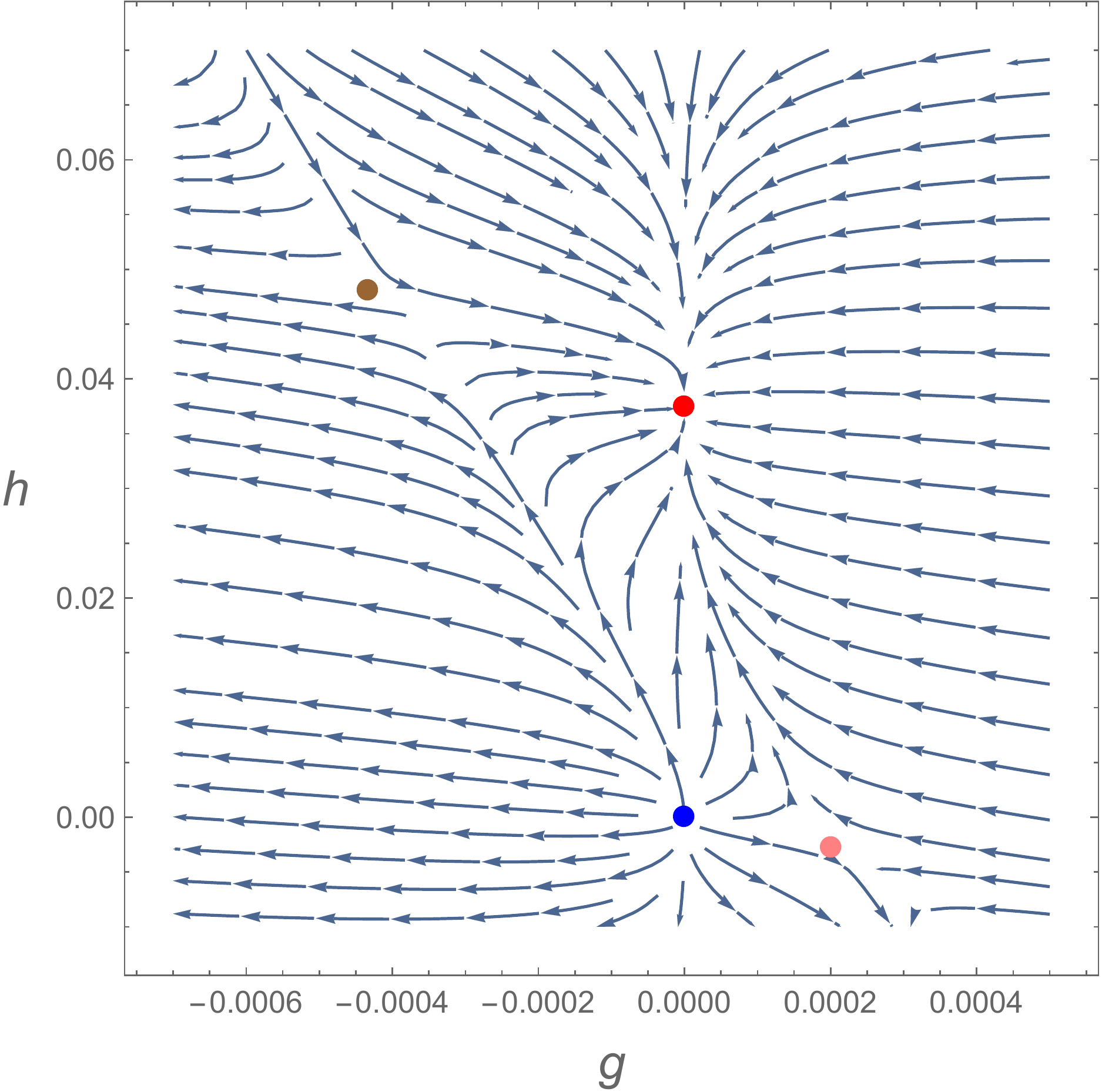}
\caption{Phase diagrams for theories with $\Box^2$ type kinetic term for $\epsilon=0.1$. Left panel: for $n=2$ one has the Gaussian and the generalized Wilson-Fisher fixed point with $\phi^4$ interaction. Right panel: for $n=3$ one has four fixed points, a Gaussian fixed point, a fixed point with a derivative interaction $\phi^2 (\partial\phi)^2$, and two fixed points with a mixture of $\phi^2(\partial\phi)^2$ and $\phi^6$ interactions. Both figures which show fixed points with $\mathbb{Z}_2$ symmetry qualitatively represent the phase diagram for all $n$ at this order of approximation} \label{fig}
\end{figure}

Similarly, one can calculate the critical exponents $\tilde\omega_i$ corresponding to derivative operators. The first and only nonzero exponent that is not affected by mixing with potential operators is 
\be 
\tilde\omega_{m-1} = \frac{(2m)!}{(m+1)!}h.
\ee
For $m=1$, the phase diagram of the flow equations \eqref{bv} and \eqref{bz} is depicted in Fig.\ref{fig} in the two-dimensional space of the dimensionless couplings $g,h$ (right panel). 
Four fixed points can be identified in the case of $n=3$, which interestingly include no fixed point with a pure $\phi^6$ interaction. Instead, there are two fixed points with a mixture of $\phi^2(\partial\phi)^2$ and $\phi^6$ interactions and a fixed point with a pure $\phi^2(\partial\phi)^2$ interaction, which are given at order $\epsilon$ by the critical couplings
\be 
\ba{lll}
g = 0 &\quad  g = \frac{\left(3 \sqrt{138}-13\right) \epsilon }{22200} &\quad  g = -\frac{\left(13+3 \sqrt{138}\right) \epsilon }{22200} \\[2mm]
h =  \frac{3\epsilon}{8} &\quad  h= \frac{\left(42-4 \sqrt{138}\right) \epsilon}{185}  &\quad  h= \frac{2\left(21+2 \sqrt{138}\right) \epsilon}{185} \,.
\ea
\ee 
This pattern extends to any $m$. 
The first of these non trivial fixed points is perhaps also interesting in the sense that it is infrared attractive. The critical coupling can be easily extended to all $m$
\be
h=\frac{2m(m+1)(2m+1)}{2+7m(m+1)} \,\frac{(m+1)!m!^2}{(2m)!^2}\,\epsilon \, .
\ee
It is then straightforward to calculate at this fixed point the critical exponents \eqref{eta2} and \eqref{cad2} in terms of $\epsilon$.
\be   \label{eta2ep}
\eta = \frac{\Gamma(\delta_c)}{\Gamma(2\!+\!\delta_c)}\,\frac{2(m\!+\!1)^2(2m\!+\!1)^2}{(3m\!+\!1)(2\!+\!7m(m\!+\!1))^2}\frac{(m\!+\!1)!^2m!^4}{(2m)!^3}\epsilon^2
\ee
Also from an analysis of the stability matrix, one can calculate the critical exponents 
\be \label{cad2ep}
\tilde\gamma_i = \frac{i!}{(i\!-\!m\!-\!1)!}\;\frac{2m(2m+1)}{2+7m(m+1)} \,\frac{m!}{(2m)!}\,\epsilon \,.
\ee 
At this fixed point and at order $\epsilon$ the stability matrix is block lower-triangular where each block consists of operators of the same number of derivatives, therefore the validity range of the above critical exponents extends to infinity. 
Furthermore, in this case it is similarly easy to calculate the critical exponents for the derivative operators $\phi^i(\partial\phi)^2$ which is again valid for all $i$
{\setlength\arraycolsep{2pt}
\bea
\tilde\omega_i &=& i! \left[\frac{3m+2}{2m+1}\frac{m+1}{(i\!-\!m\!)!} + \frac{1}{(i\!-\!m\!-\!1)!} \right. \\
&& \left. \hspace{15mm} +\frac{m(m+1)}{(i-m+1)!}\right]\frac{2m(2m+1)}{2+7m(m+1)} \frac{m!}{(2m)!}\epsilon .\nn
\eea}%
The phase diagram of the $n=3$ case is compared with that of $n=2$ in Fig.\eqref{fig} (left panel), where $g, h$ are respectively the coefficients of the operators $\varphi^{4}$ in $v(\varphi)$ and
$\varphi (\partial\varphi)^2$ in $z(\varphi)$. 
If one includes higher order corrections for the beta functions the $\phi^{4}$ fixed point persists, while in principle other fixed points may arise which have derivative interactions. However, as discussed in Section \eqref{sec.general}, these have an odd $\mathbb{Z}_2$ parity and can be consistently ignored.
%
%

\noindent
{\bf \it CFT approach.}
The same line of reasoning as that discussed towards the end of Section \eqref{sec.type1} can be followed also in this case to extract critical properties. With a knowledge of the non-trivial structure of the critical theory encoded in the Landau-Ginzburg Lagrangian~\eqref{LGnew} with $Z(\phi)=h \phi^{2m}$ and $V(\phi)=g \phi^{2(2m+1)}$ one can write down the equation of motion
\be
0=\square^2\phi\! +\! 2(2m\!+\!1) g\phi^{4m+\!1}\!-\! mh\phi^{2m\!-\!1}(\partial\phi)^2\! -\! h\phi^{2m}\square\phi.
\ee 
This can be used, for instance, to re-derive the leading corrections to the anomalous dimensions of $\phi^i$.
For $\gamma_1$ one acts directly with Laplacians on the two point function \nolinebreak as
\be 
\square^2_x\square^2_y\langle\phi_x \phi_y\rangle \stackrel{\mathrm{LO}}{=} - 2^9\gamma_1\,c\,|x-y|^{-\frac{2}{m}-8}\prod_{i=0}^3 (i+1/m),
\ee
where $c^m = 4\Gamma(\delta_c)^{m}/(4\pi)^{2m+1}$. This can be compared with the same quantity evaluated using the SDE
\bea
\!\!\langle \square^2_x\phi_x \square^2_y\phi_y\rangle &\stackrel{\mathrm{LO}}{=}& 4(2m\!+\!1)^2  (4m+1)!g^2 c^{4m+1}|x\!-\!y|^{-\frac{2}{m}-8} \nn\\
&-& 8(2m\!+\!1)!m^{-2} h^2 c^{2m+1}|x\!-\!y|^{-\frac{2}{m}-8}.
\eea
After rescaling the couplings $g,h$ in accord with our RG conventions \eqref{res2}, one correctly reproduces $\gamma_1$ which is one-half $\eta$ in \eqref{eta2}. 
In a similar way, for the anomalous dimensions $\gamma_i$ one can compare the action of $\Box^2$ on the three point function $\left\langle \phi_x\phi^i_y\phi^{i+1}_z\right\rangle$, evaluated both directly and by using the SDE. 
Comparing the two 
at leading order, one gets in terms of the rescaled couplings the recursion relation 
\be 
\gamma_{i+1} - \gamma_i = \binom{i}{m} \frac{(2m)!}{(m+1)!} h + 4\binom{i}{2m} \frac{(4m+1)!}{(2m)!} g,
\ee
which can be solved upon imposing the boundary condition $\gamma_1=0$. 
This leads to complete agreement with the quantities obtained within the RG method in Eq.~\eqref{cad2}.

\section{Conclusion and Outlook}

We have explored a class of multicritical scalar theories with higher-derivative kinetic term. As a tool to analyze such models we employed RG which relies neither on unitarity nor conformal invariance. One of the advantages of such an approach has been to allow to correctly identify scale invariant deformations of higher-derivative free CFTs.
In particular, for the class of theories where $k$ and $n-1$ have a common divisor the fixed points of the RG flow correspond to critical theories with derivative interactions. This causes the potential approximation to break down and makes the inclusion of derivative operators unavoidable, unveiling a novel pattern which was missed in previous CFT-based analyses~\cite{Gliozzi:2016ysv,Gliozzi:2017hni}. 
%
As an illustrative example we have analyzed in detail theories with $\Box^2$ kinetic terms and odd values of $n$, 
which correspond to the second type, and shown explicitly that pure potential deformations of  higher-derivative free CFTs are not scale invariant. In these models, we have identified instead, among others, a pure derivative scale invariant deformation which is also infrared attractive.

We have also confirmed most of our RG results, by making use of the SDE and assuming conformal symmetry. Our findings for these non-unitary critical theories therefore provide evidence for conformal invariance, at least at the leading non-trivial perturbative order in $\epsilon$. 
It would still be interesting to investigate in particular theories of the second type with other approaches which rely on conformal symmetry, such as that based on the analytic structure of conformal blocks, to reproduce and perhaps extend the results presented here.

In this work we concentrated on integer values of the parameter $n$, which correspond to even potentials $\phi^{2n}$ or more generally $\mathbb{Z}_2$ symmetric fixed points, but our method can be easily extended to critical models with odd potentials as well. 
Among other critical higher derivative theories one can think of those characterized by shift symmetry, and even consider multiple scalar fields with global symmetries, 
e.g. $O(N)$ models. It would be interesting to carry on an analysis with both  RG and, whenever applicable, CFT methods at perturbative level,
and finally  to study all such kind of theories non perturbatively, using functional RG methods and, despite their non unitary nature, with the non perturbative conformal bootstrap as well.
\paragraph*{Acknowledgments}

We thank A. Petkou for discussions.



\end{document}